\documentclass{ws-p8-50x6-00}

\newcommand{\PsiDY} {J/$\psi$/DY}
\newcommand{\Et} {E$_{\rm T}$}
\newcommand{\Ezdc} {E$_{\rm ZDC}$}
\newcommand{\etamx} {$\eta_{\rm max}$}
\newcommand{\dNdeta} {dN$_{\rm ch}$/d$\eta$}
\newcommand{\bce} {\begin{center}}
\newcommand{\ece} {\end{center}}
\begin{document}

%\title{}
%\author{}
%\address{}
%\maketitle
\begin{center}
{\bf CENTRALITY BEHAVIOUR OF J/$\psi$ PRODUCTION IN NA50} \\
\vspace{3mm}
{\bf Presented by S\'ERGIO RAMOS$^{6,b}$} \\
\vspace{3mm}
{\bf NA50 Collaboration}

{\small
M.C.~Abreu$^{6,a}$,
B.~Alessandro$^{10}$,
C.~Alexa$^{3}$,
R.~Arnaldi$^{10}$,
%\vspace{-0.5 mm}
M.~Atayan$^{12}$,
C.~Baglin$^{1}$,
A.~Baldit$^{2}$,
M.~Bedjidian$^{11}$,
S.~Beol\`e$^{10}$,
%\vspace{-0.5 mm}
V.~Boldea$^{3}$,
P.~Bordalo$^{6,b}$,
S.R.Borenstein$^{9,c}$,
G.~Borges$^{6}$,
A.~Bussi\`ere$^{1}$,
L.~Capelli$^{11}$,
%\vspace{-0.5 mm}
J.~Castor$^{2}$,
C.Castanier$^{2}$,
B.~Chaurand$^{9}$,
B.~Cheynis$^{11}$,
E.~Chiavassa$^{10}$,
%\vspace{-0.5 mm}
C.~Cical\`o$^{4}$,
T.~Claudino$^{6}$,
M.P.~Comets$^{8}$,
N.~Constans$^{9}$,
S.~Constantinescu$^{3}$,
%\vspace{-0.5 mm}
P.~Cortese$^{10}$,
J.~Cruz$^{6}$,
N.~De Marco$^{10}$,
A.~De Falco$^{4}$,
%\vspace{-0.5 mm}
G.~Dellacasa$^{10,d}$,
A.~Devaux$^{2}$,
S.~Dita$^{3}$,
O.~Drapier$^{11}$,
%\vspace{-0.5 mm}
B.~Espagnon$^{2}$,
J.~Fargeix$^{2}$,
P.~Force$^{2}$,
M.~Gallio$^{10}$,
Y.K.~Gavrilov$^{7}$,
%\vspace{-0.5 mm}
C.~Gerschel$^{8}$,
P.~Giubellino$^{10}$,
M.B.~Golubeva$^{7}$,
M.~Gonin$^{9}$,
%\vspace{-0.5 mm}
A.A.~Grigorian$^{12}$,
S.~Grigorian$^{12}$,
J.Y.~Grossiord$^{11}$,
F.F.~Guber$^{7}$,
A.~Guichard$^{11}$,
%\vspace{-0.5 mm}
H.~Gulkanyan$^{12}$,
R.~Hakobyan$^{12}$,
R.~Haroutunian$^{11}$,
M.~Idzik$^{10,e}$,
%\vspace{-0.5 mm}
D.~Jouan$^{8}$,
T.L.~Karavitcheva$^{7}$,
L.~Kluberg$^{9}$,
A.B.~Kurepin$^{7}$,
%\vspace{-0.5 mm}
Y.~Le~Bornec$^{8}$,
C.~Louren\c co$^{5}$,
P.~Macciotta$^{4}$,
M.~Mac~Cormick$^{8}$,
%\vspace{-0.5 mm}
A.~Marzari-Chiesa$^{10}$,
M.~Masera$^{10}$,
A.~Masoni$^{4}$,
%S.~Mehrabyan$^{12}$,
%\vspace{-0.5 mm}
M.~Monteno$^{10}$,
A.~Musso$^{10}$,
P.~Petiau$^{9}$,
A.~Piccotti$^{10}$,
J.R.~Pizzi$^{11}$,
%\vspace{-0.5 mm}
W.L.~Prado da Silva$^{10,f}$,
F.~Prino$^{10}$,
G.~Puddu$^{4}$,
C.~Quintans$^{6}$,
%\vspace{-0.5 mm}
S.~Ramos$^{6,b}$,
L.~Ramello$^{10,d}$,
P.~Rato Mendes$^{6}$,
L.~Riccati$^{10}$,
%\vspace{-0.5 mm}
A.~Romana$^{9}$,
H.~Santos$^{6}$,
P.~Saturnini$^{2}$,
E.~Scalas$^{10,d}$,
E.~Scomparin$^{10}$,
S.~Serci$^{4}$,
%\vspace{-0.5 mm}
R.~Shahoyan$^{6,g}$,
F.~Sigaudo$^{10}$,
S.~Silva$^{6}$,
M.~Sitta$^{10,d}$,
P.~Sonderegger$^{5,b}$,
%\vspace{-0.5 mm}
X.~Tarrago$^{8}$,
N.S.~Topilskaya$^{7}$,
G.L.~Usai$^{4}$,
E.~Vercellin$^{10}$,
L.~Villatte$^{8}$,
N.~Willis$^{8}$
}
\end{center}
\vskip 2mm
{\footnotesize
\noindent
$^{1}$ LAPP, CNRS-IN2P3, Annecy-le-Vieux,  France;~
$^{2}$ LPC, Universit\'e Blaise Pascal and CNRS-IN2P3, Aubi\`ere, France;~
$^{3}$ IFA, Bucharest, Romania;~
$^{4}$ Universit\`a di Cagliari/INFN, Cagliari, Italy;~
$^{5}$ CERN, Geneva, Switzerland;~
$^{6}$ LIP, Lisbon, Portugal;~
$^{7}$ INR, Moscow, Russia;~
$^{8}$ IPN, Univ. de Paris-Sud and CNRS-IN2P3, Orsay, France;~
$^{9}$ LPNHE, Ecole Polytechnique and CNRS-IN2P3, Palaiseau, France;~
$^{10}$ Universit\`a di Torino/INFN, Torino, Italy;~
$^{11}$ IPN, Universit\'e Claude Bernard Lyon-I and CNRS-IN2P3, Villeurbanne, France;~
$^{12}$ YerPhI, Yerevan, Armenia.
\vskip 1.5mm
\noindent
$^{a}$ also at FCT, Universidade de Algarve, Faro, Portugal;~
$^{b}$ also at IST, Universidade T\'ecnica de Lisboa, Lisbon, Portugal;~
$^{c}$ on leave of absence from York College CUNY;~
$^{d}$ Universit\`a del Piemonte Orientale, Alessandria and INFN-Torino, Italy;~
$^{e}$ also at Fac. Physics and Nuclear Techniques, Univ. Mining and Metallurgy,
Cracow, Poland;~
$^{f}$ now at UERJ, Rio de Janeiro, Brazil;~
$^{g}$ on leave of absence of YerPhI, Yerevan, Armenia.
}
\vskip 5mm

\abstracts{The J/$\psi$ production in 158 A GeV Pb-Pb interactions is studied,
in the dimuon decay channel, as a function of centrality, as measured with the
electromagnetic or with the very forward calorimeters. After a first
sharp variation at mid centrality, both patterns continue to fall down and
exhibit a curvature change at high centrality values. This trend excludes any
conventional hadronic model and is in agreement with a deconfined quark-gluon
phase scenario. We report also preliminary results on the measured charged
multiplicity, as given by a dedicated detector.}

\vspace{-3mm}
\section{Introduction}
\vspace{-1.5mm}
NA50 is a dimuon experiment searching for specific signals of deconfinement, 
namely the predicted suppression of charmonia production. Indeed, it has been
predicted\cite{MatSatz} that the $c\bar{c}$ bound states are prevented to be
formed by the colour screening potential in the very dense medium undergoing
a phase transition to a deconfined medium of quarks and gluons.

\vspace{-3mm}
\section{Experimental setup and data selection}
\vspace{-1.5mm}
The NA50 apparatus\cite{NA50_410_327} mainly consists of a muon spectrometer, a segmented
active target and three independent centrality detectors: an electromagnetic
calorimeter which measures the neutral transverse energy (\Et) produced in the
interaction, a zero degree calorimeter measuring the very forward hadronic
energy (\Ezdc) of the spectator nucleons and a silicon strip multiplicity
detector.

In this analysis we use data taken in 1996 and 1998 with a lead beam impinging
on a Pb target, as well as the new p-A data of 1998-2000 (A $\equiv$ Be, Cu, Al,
Ag, W, Pb) taken at two different beam intensities.

The kinematical domain used for dimuon detection, $2.92~\le~y_{lab}~\le~3.92$
%(i.e., $0~\le~y_{cms}\le~1$)
and $|cos~\theta_{CS}|~<~0.5$ leads, in the mass region of inte\-rest, to
acceptances of the order of 15\%. The centrality detectors cover the
following rapidity domains: e.m. calorimeter, $1.1<\eta_{lab}<2.3$~; zero
degree calorimeter, $\eta_{lab}>6.3$~;  multiplicity detector,
$0.82<\eta_{lab}<4.18$~.

The J/$\psi$ is detected via its decay into muon pairs. Combinatorial
background, due to $\pi$ and $K$ decays, is estimated from like-sign
pairs\cite{NA50_410_327}, using $N_{BG}=2 \sqrt{N^{++}N^{--}}$~.
The muon pairs selected for the analyses satisfy the standard NA50
criteria\cite{NA50_410_337}.

\vspace{-3mm}
\section{J/$\psi$ production analyses}
\vspace{-1.5mm}
A study of Drell-Yan behaviour, from p-p and several p-A systems to S-U and
Pb-Pb, proves that its cross-section behaves normally and is proportional to
the number of elementary nucleon-nucleon collisions\cite{ichep98}
%(i.e., the product A$\cdot$B)~
. Thus, it can be used as a J/$\psi$ reference.

A J/$\psi$ systematic study is performed\cite{NA50_450} using all our
previous data from lighter systems, ranging from p-p to S-U
(NA38\cite{NA38_444,NA38_449,NA38_466} and NA51\cite{NA51}). The data are
described by a simple exponential parametrization:
$B_{\mu\mu} \sigma^{\psi}/\sigma^{DY}~\propto~e^{- \rho L \sigma_{abs}}$
(L being the path length of the pre-resonant state in nuclear matter),
giving an absorption value of the $c\bar{c}g$ state in nuclear matter of
$\sigma_{abs} = 5.8 \pm 0.6$ mb (full calculation gives $6.4 \pm 0.5$ mb).
It is the normal J/$\psi$ suppression.

\begin{figure}[!ht]
\vspace*{-2mm}
\begin{center}
 \begin{minipage}[c]{.32\linewidth}
  \includegraphics[width=\linewidth]{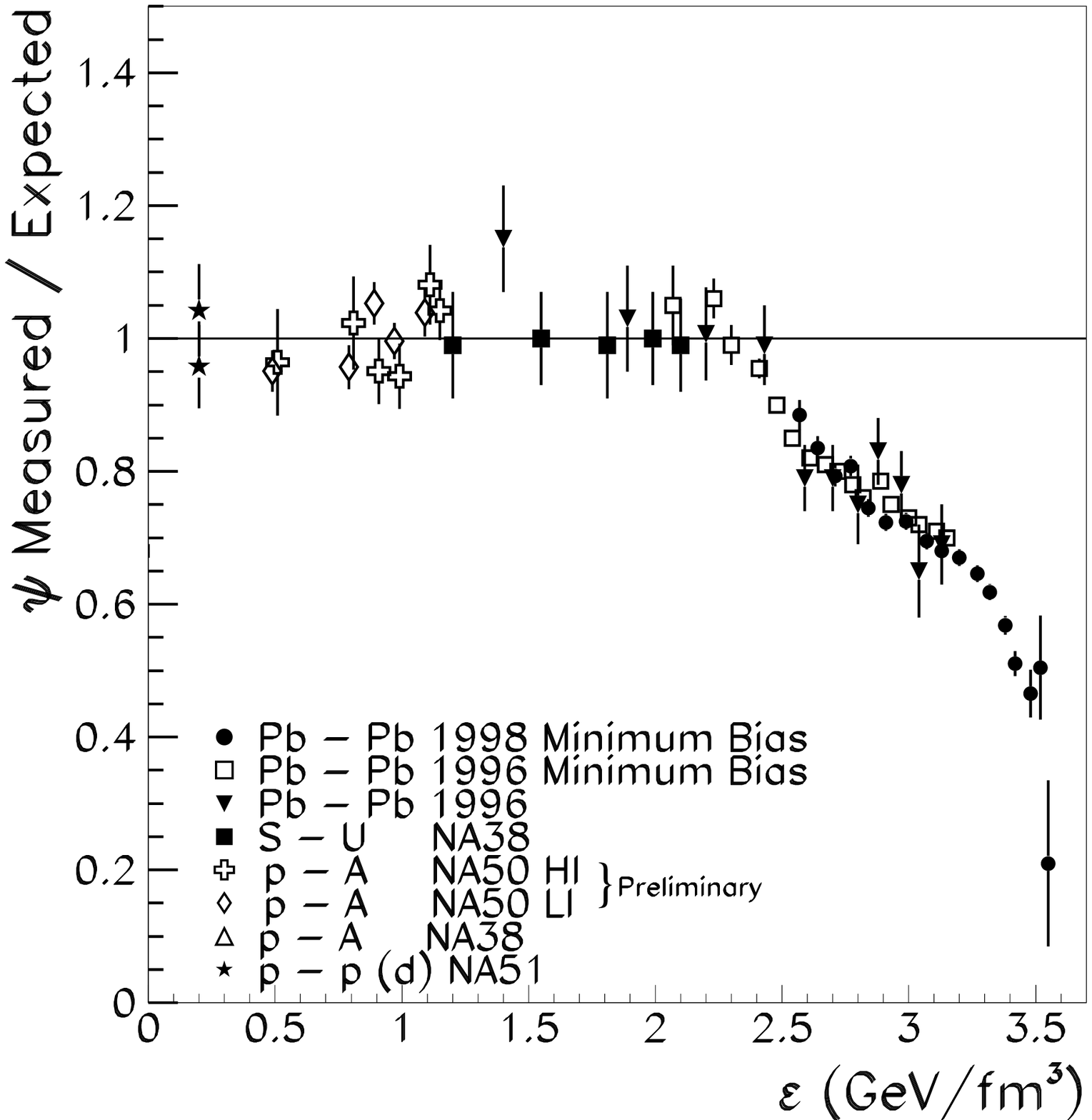}
\vspace*{-4mm}
\caption{J/$\psi$ production normalized to the expected yield (
%our nuclear
absorption fit from p-p to S-U) as a function of $\epsilon$}
\label{fig:psieps}
 \end{minipage} \hfill
 \begin{minipage}[c]{.32\linewidth}
  \includegraphics[width=\linewidth]{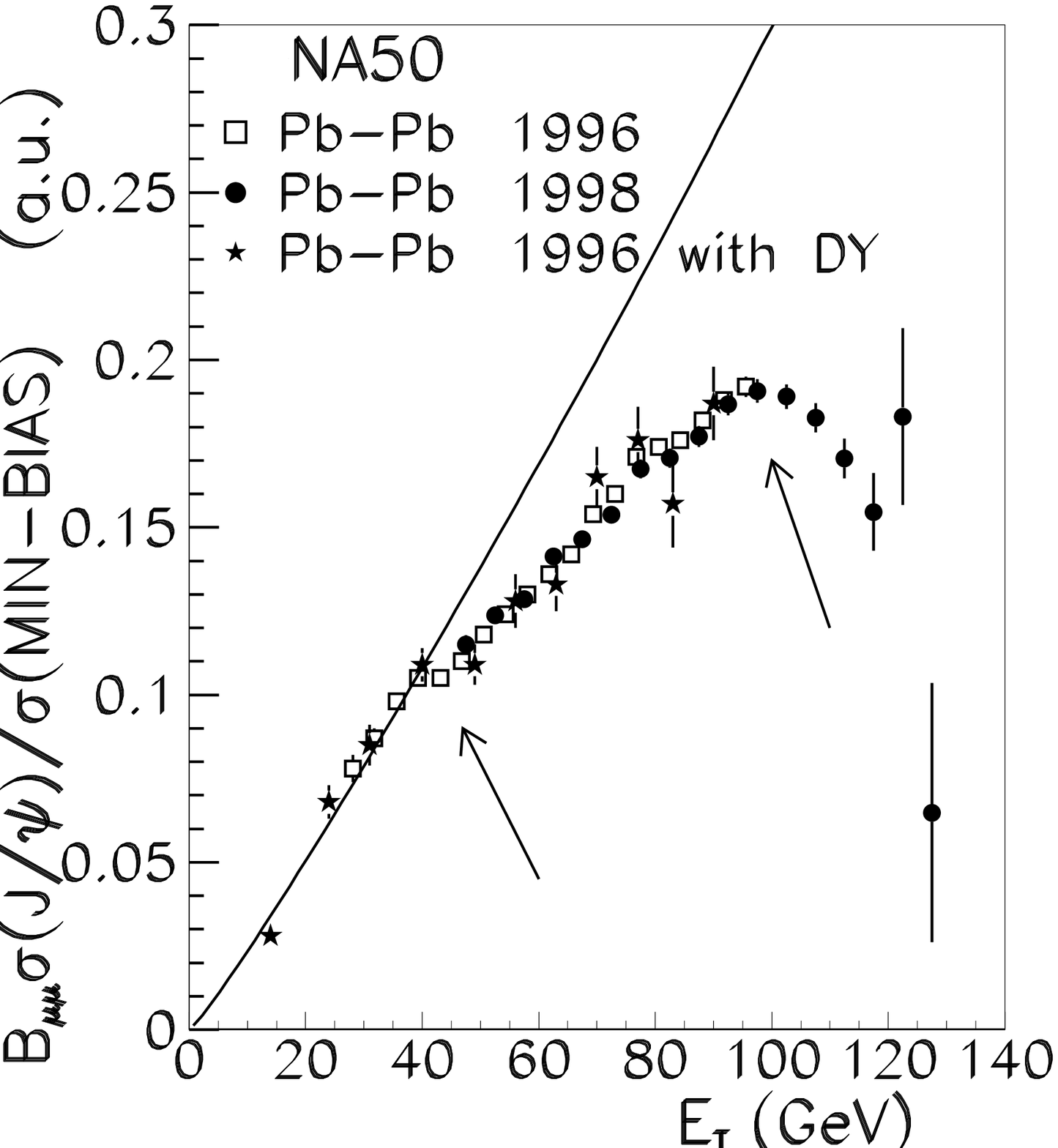}
\vspace*{-7mm}
\caption{J/$\psi$/Minimum Bias~ as a function of \Et.
The curve stands for our absorption fit}
\label{fig:psimb}
\end{minipage} \hfill
 \begin{minipage}[c]{.32\linewidth}
\vspace*{1mm}
  \includegraphics[width=\linewidth]{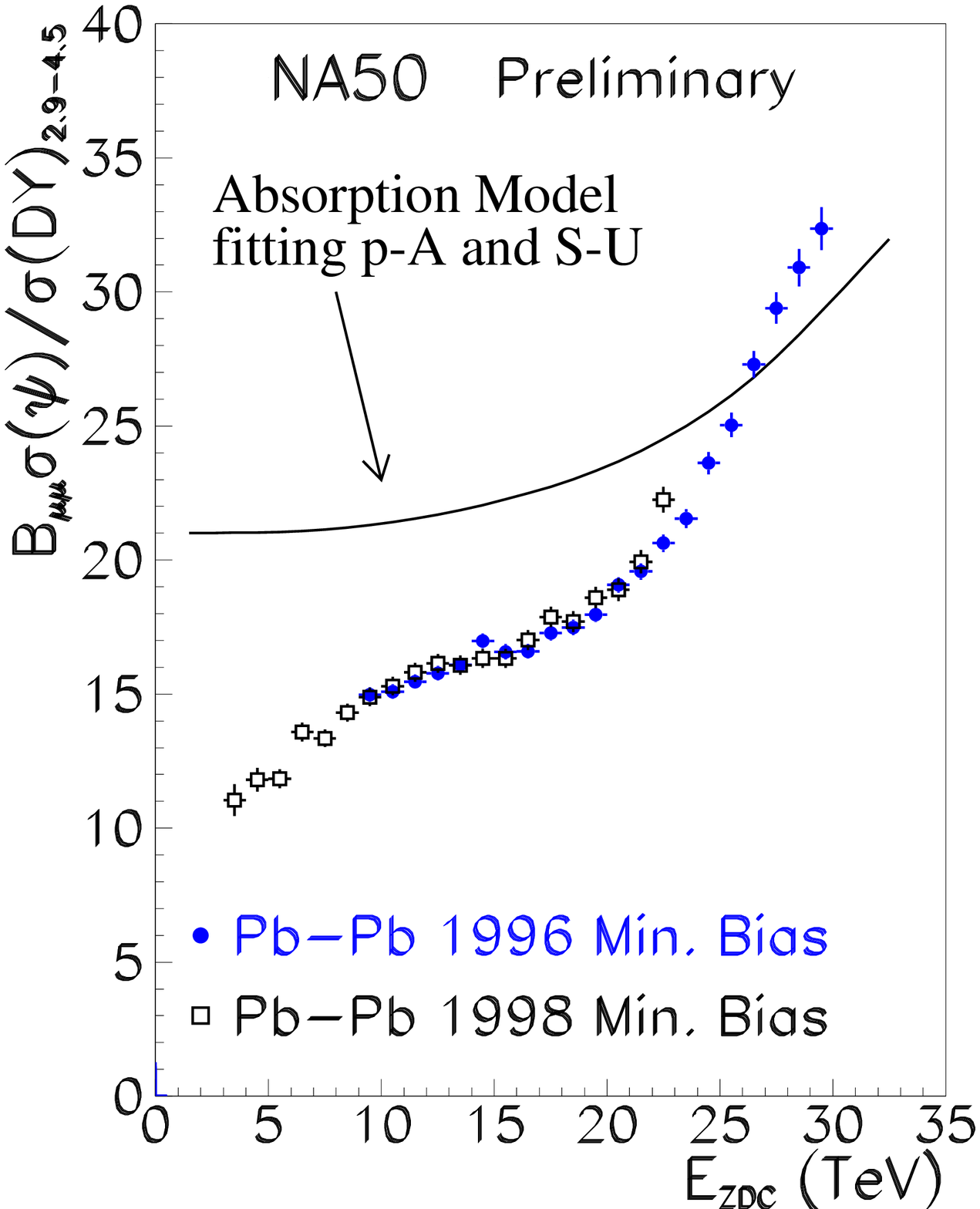}
\vspace*{-7mm}
\caption{\PsiDY~ as a function of \Ezdc\hspace{4 cm}}
\label{fig:psizdc}
\end{minipage}
\end{center}
\vspace*{-4mm}
\end{figure}

In Fig.~\ref{fig:psieps} we show the J/$\psi$ production pattern, normalized
to our absorption fit from p-p to S-U, as a function of the energy density
$\epsilon$ (evaluated by the Bjorken's model), putting together all our
systems, from p-p to Pb-Pb\cite{NA50_477}. Whereas the more peripheral
Pb-Pb points lie on the absorption curve, the more central ones show a sudden
20\% drop at 2.3 GeV/fm$^3$ followed by an inflection point at 3.1 GeV/fm$^3$.
It is the %so-called 
anomalous J/$\psi$ suppression.

The J/$\psi$ production is also studied, normalized to the minimum bias
cross-section, as a function of \Et~(Fig.~\ref{fig:psimb}). Once again, a
clear departure from our absorption fit at 40 GeV (corresponding to L = 8 fm)
and an inflection point at 90 GeV followed by a steady steep decrease are
observed, reinforcing the idea of the onset of another $J/\psi$ suppression
mechanism.

Finally, \PsiDY~ is studied as a function of an independent centrality
variable, as given by a different detector, the forward hadronic
calorimeter (Fig.~\ref{fig:psizdc}). One observes, again, the stepwise
J/$\psi$ suppression pattern.

\begin{figure}[!ht]
%\vspace*{-12mm}
\begin{center}
 \begin{minipage}[c]{.48\linewidth}
%\vspace*{-10mm}
  \includegraphics[width=\linewidth]{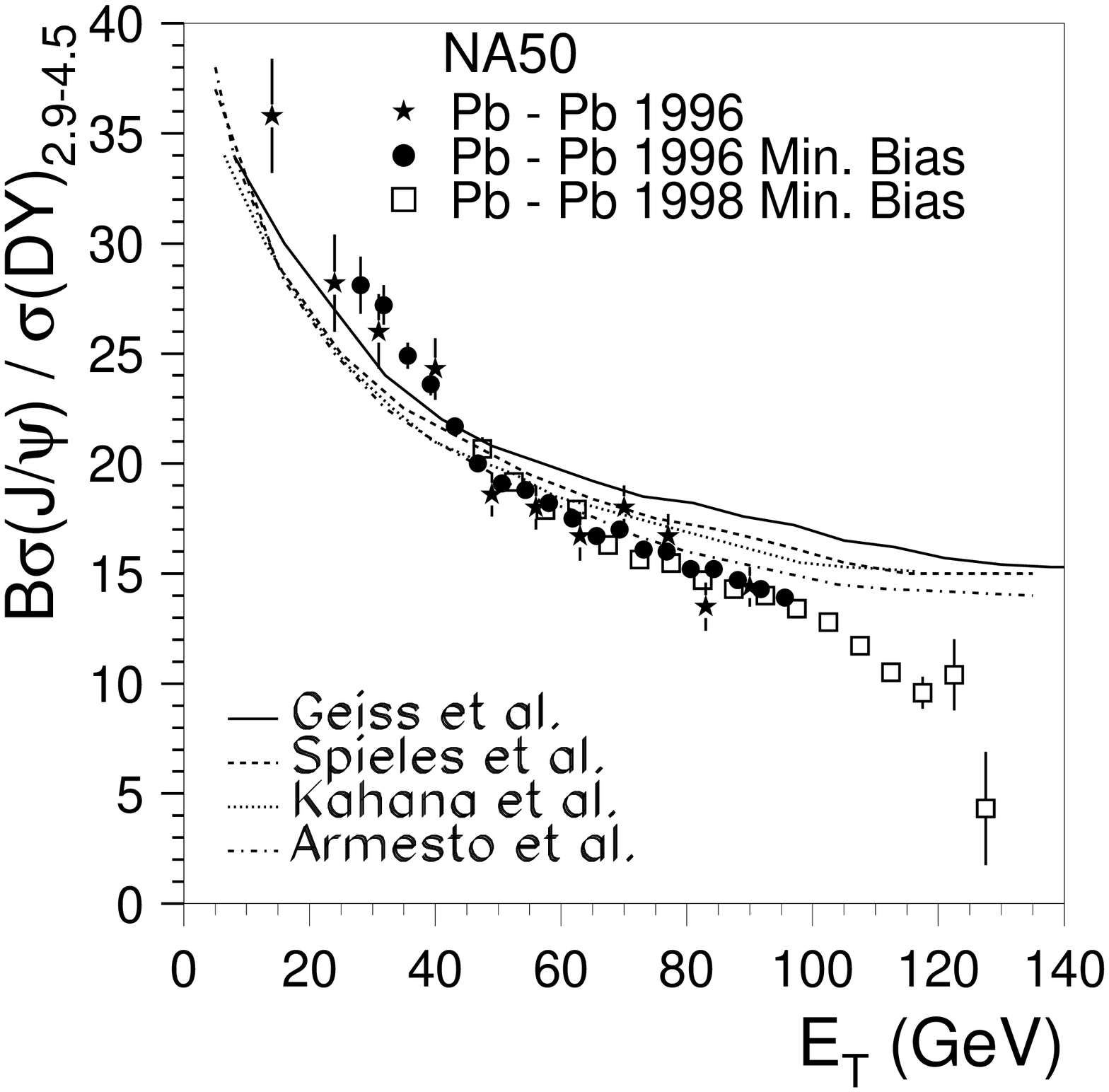}
\vspace*{-10mm}
\caption{The \PsiDY~ stepwise pattern compared with conventional
(hadronic) models based on comovers}
\label{fig:modcom}
\end{minipage} \hfill
 \begin{minipage}[c]{.48\linewidth}
  \includegraphics[width=\linewidth]{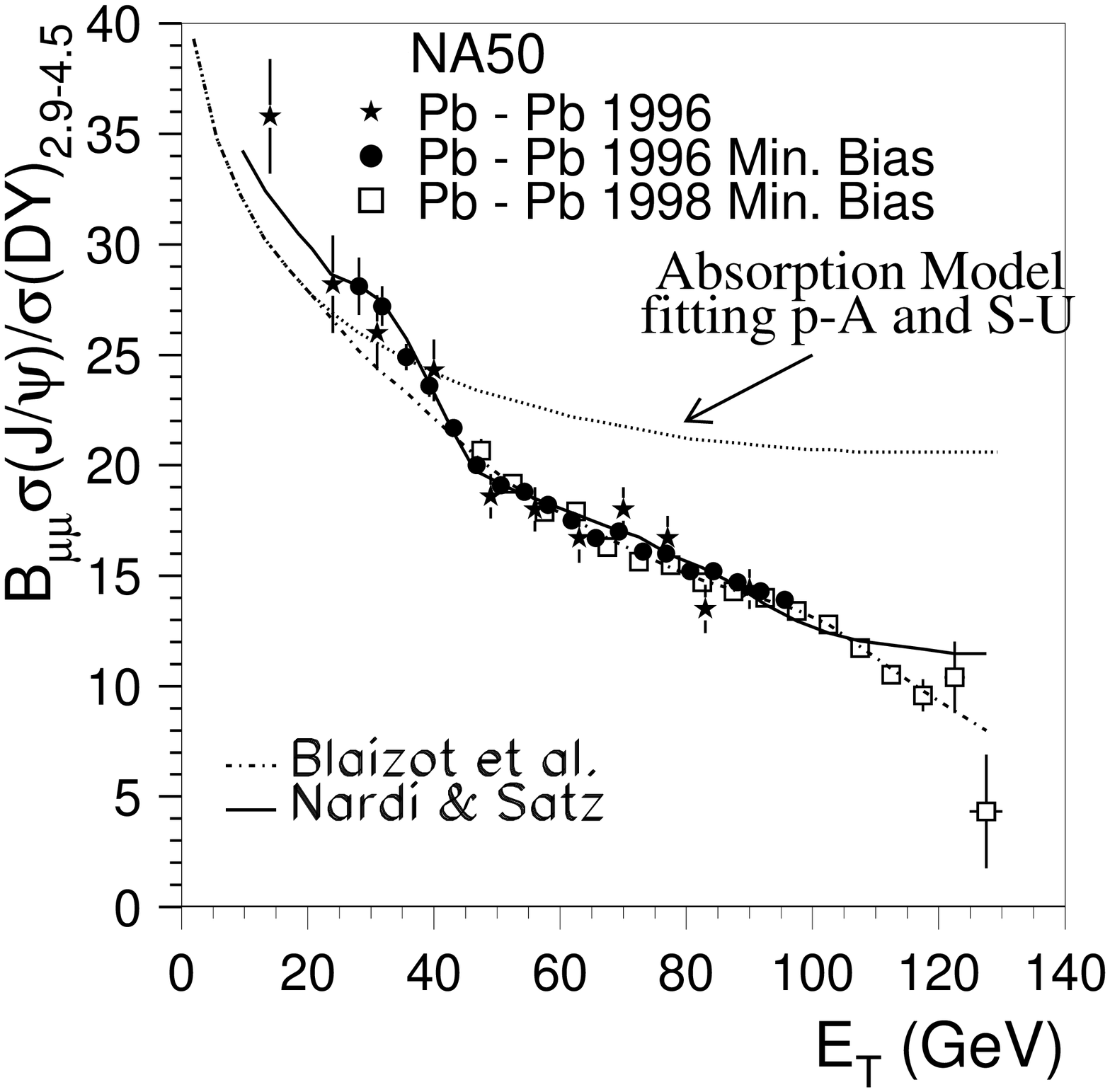}
\vspace*{-10mm}
\caption{The \PsiDY~ stepwise pattern compared with models
assuming deconfinement}
\label{fig:modqgp}
\end{minipage}
\end{center}
\vspace*{-8mm}
\end{figure}

Figures \ref{fig:modcom} and \ref{fig:modqgp} show the comparision of J/$\psi$
production with some mo\-dels: on the left, models\cite{rqmd,kahana,armesto,capella}
assume that J/$\psi$ is absorbed by comovers; on the right,
models\cite{satz,blaizot} assume quark-gluon deconfinement.
One sees that the J/$\psi$ stepwise suppression pattern rules out any of
these hadronic models, and may be explained in the framework of quark-gluon
deconfinement, as melting of charmonium states.

\vspace{-3mm}
\section{Multiplicity distributions}
\vspace{-1.5mm}
Charged particle multiplicities \dNdeta~ were also studied, using the
multiplicity detector\cite{NA50_410_327}, which is approximately centered at
mid-rapidity. For this analysis, 2.2 $\eta$ units were used:
$1.89 \le \eta_{lab} \le 4.18$~.

Events were subdivided in 7 contiguous \Ezdc~ regions, as defined by our forward
hadronic calorimeter (ZDC). This allows the definition of centrality classes,
%corresponding to 5\% of the total inelastic cross-section
of about 5\% $\sigma_{\rm inel}$ each.

\begin{figure}[!ht]
\vspace*{-13.5mm}
\begin{center}
 \begin{minipage}[c]{.48\linewidth}
%\vspace*{-10mm}
\vspace*{7mm}
  \includegraphics[width=\linewidth]{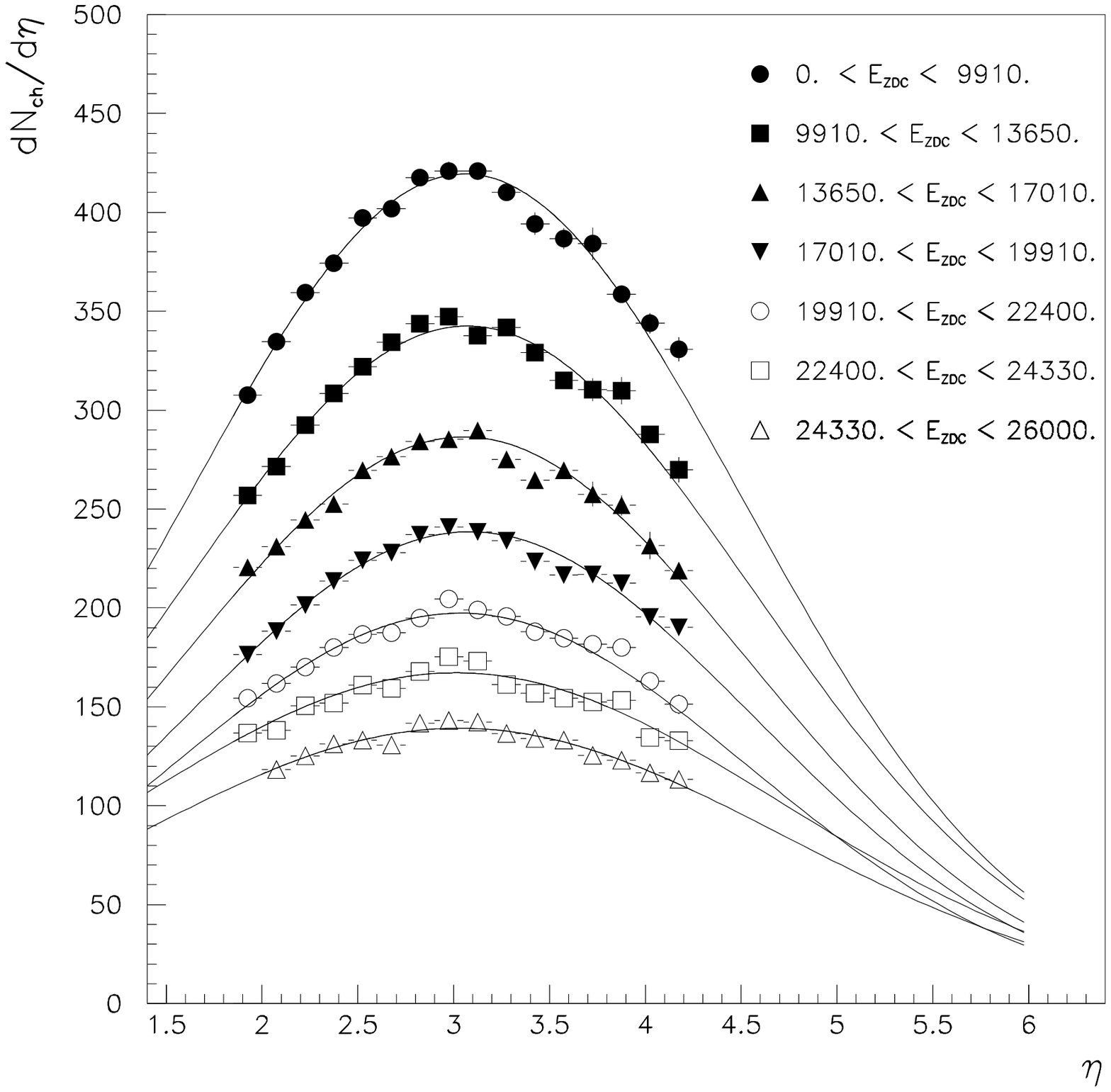}
\vspace*{-10mm}
\caption{Pseudorapidity peak positions, \etamx, as a function of
the forward energy \Ezdc~}
\label{fig:mult1}
\end{minipage} \hfill
 \begin{minipage}[c]{.48\linewidth}
\vspace*{5mm}
  \includegraphics[width=\linewidth]{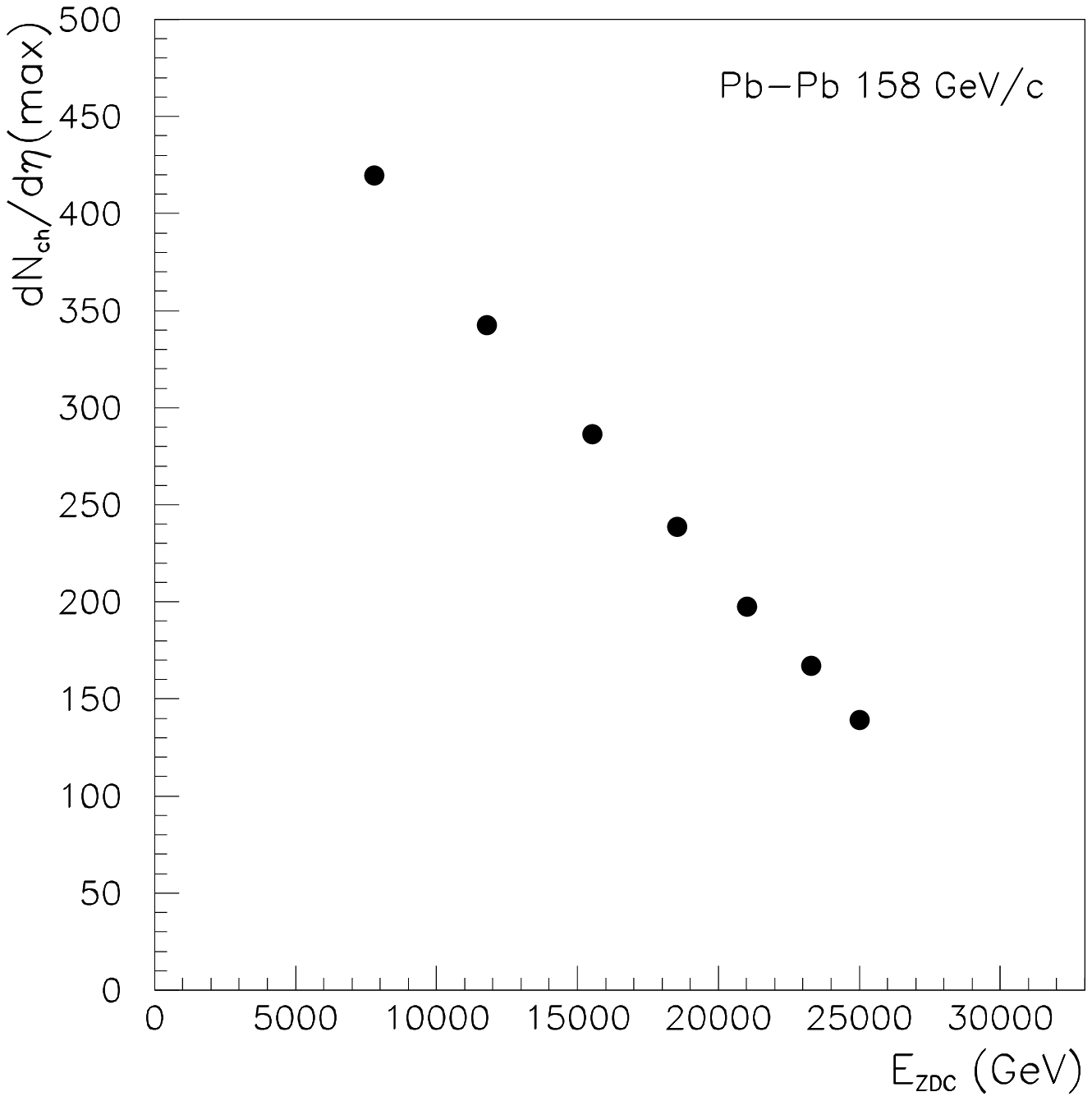}
\vspace*{-10mm}
\caption{Number of charged particles at peak, \dNdeta$|_{max}$,
as a function of \Ezdc~}
\label{fig:mult2}
\end{minipage}
\end{center}
\vspace*{-6mm}
\end{figure}

Multiplicity distributions were fitted with gaussians (see Fig.~\ref{fig:mult1}),
from which we evaluate the peak positions, $\eta_{\rm max}$, and obtain the
maximum values of the charged particle densities, \dNdeta$|_{\rm max}$~.

For all centrality classes, peak positions are compatible with a constant value
of 3.10, in agreement with VENUS 4.12 Monte-Carlo\cite{venus} expectations.

Widths of \dNdeta~ gaussian fits slightly decrease with centrality (from 1.6
to 1.4). This can be attributed to the higher degree of stopping in central
collisions as compared to peripheral ones.

The pseudorapidity density of N$_{\rm ch}$~, that is, the \dNdeta$|_{max}$
values show a linear behaviour with centrality, evaluated by the forward
energy measured with ZDC calorimeter (Fig.~\ref{fig:mult2}).

\vspace{-3mm}
\section{Conclusions}
\vspace{-1.5mm}
Concerning J/$\psi$ production, three different types of analysis have been
performed, namely J/$\psi$ cross-section normalized to the Drell-Yan one as a
function of two independent centrality variables (the transverse energy
measured with the electromagnetic calorimeter, and the very forward energy
evaluated with the zero degree calorimeter) and J/$\psi$ cross-section
normalized to the minimum bias one, as a function of \Et~. They all agree
with the observed stepwise pattern of the anomalous J/$\psi$ suppression:
a drop of about 20\% and an inflection point followed by a steady steep
decrease. Hadronic models with comovers can not describe our data (systems
ranging from p-p and several p-A to S-U and Pb-Pb), which can be interpreted
as melting of charmonium states in a deconfined medium.
%($\chi_c$, $J/\psi$) as predicted by Quark-Gluon Deconfinement.

Concerning multiplicity studies, one notices that the $\eta$ peak position
stays constant for all centrality classes (as predicted by VENUS Monte-Carlo)
and observes an approximately linear increase of \dNdeta~ with centrality
(\Ezdc).

\end{document}